\documentclass{article}
\usepackage{tipa}
\usepackage{tikz}
\usepackage{environ}
\usetikzlibrary{calc} 
\usepackage{amsmath,graphicx}
\usepackage[preprint]{spconf}
\usepackage{graphicx}
\usepackage{lipsum,adjustbox}
\usepackage{amssymb,amsmath,bm}
\usepackage[bottom]{footmisc}
\usepackage{dblfloatfix}
\usepackage{textcomp}
\usepackage{microtype}
\usepackage{booktabs}
\usepackage{multirow}
\usepackage{cite}
\usepackage{placeins}
\usepackage{float}
\usepackage{balance}
\usetikzlibrary{shapes.geometric, arrows}
\tikzstyle{process} = [rectangle, minimum width=1.25cm, minimum height=1cm, outer sep=0,text centered, draw=black, fill=white]
\tikzstyle{c} = [circle, minimum size=1.5em,inner sep=0.2,text centered, draw=black, fill=white]
\tikzstyle{arrow} = [thick,->,>=stealth]
\tikzstyle{tria} = [regular polygon, regular polygon sides=3,draw, fill=white, text width=1em,inner sep=0.3mm, outer sep=0mm,shape border rotate=-90]
\tikzstyle{tria_r} = [regular polygon, regular polygon sides=3,draw, fill=white, text width=1em,inner sep=0.3mm, outer sep=0mm,shape border rotate=90]
\title{ImportantAug: a data augmentation agent for speech}
\name{Viet Anh Trinh${^1}$, Hassan Salami Kavaki ${^1}$ and Michael I Mandel${^{1,2}}$}
\address{
  ${^1}$ The Graduate Center, CUNY, New York, USA\\
  ${^2}$ Brooklyn College, CUNY, New York, USA \\
  vtrinh@gradcenter.cuny.edu, hsalami@gradcenter.cuny.edu, mim@sci.brooklyn.cuny.edu}
\begin{document}
\ninept
\maketitle

\copyrightnotice{ \begin{tabular}[t]{@{}l@{}} To appear in {\it Proc.\ ICASSP 2022, May, 2022, Singapore.} \\
\copyright\ 2022 IEEE. Personal use of this material is permitted. Permission  from IEEE must be obtained for all other uses,  in any current  or  future \\ media,  including reprinting/republishing this material  for advertising or promotional purposes, creating new collective  works, for resale or \\ redistribution to  servers or lists,  or reuse of any copyrighted component of this work in other works. \end{tabular} } 
\begin{abstract}
We introduce ImportantAug, a technique to augment training data for speech classification and recognition models by adding noise to unimportant regions of the speech and not to important regions. Importance is predicted for each utterance by a data augmentation agent that is trained to maximize the amount of noise it adds while minimizing its impact on recognition performance. The effectiveness of our method is illustrated on version two of the Google Speech Commands (GSC) dataset. On the standard GSC test set, it achieves a 23.3\% relative error rate reduction compared to conventional noise augmentation which applies noise to speech without regard to where it might be most effective. It also provides a 25.4\% error rate reduction compared to a baseline without data augmentation. Additionally, the proposed ImportantAug outperforms the conventional noise augmentation and the baseline on two test sets with additional noise added. 
\end{abstract}
\section{Introduction}
Data augmentation techniques are used to enhance models' performance by adding additional variations to the training data. These techniques are widely applied to improve automatic speech recognition (ASR) performance \cite{ko15_interspeech,kim17_interspeech,Hannun2014DeepSS,nguyen2020improving}. In \cite{ko15_interspeech}, the authors used speed perturbation to create new speech utterances by changing the frequency components and number of time frames of speech recordings. This additional training data helped to decrease the word error rate (WER) by 3.2\% relative on Librispeech task with 960 hours Librispeech data. In \cite{kim17_interspeech}, reverberation was added to the speech to make it more realistic. Recently, a common technique is to remove or mask information in the spectrogram domain. For instance, SpecAugment \cite{park2019specaugment} removes speech information in $T$ continuous random time frames or $F$ frequency bins. At the time, this augmentation not only increased ASR accuracy, but also achieved the state-of-the-art WER on the LibriSpeech 960-hour dataset at 5.8\%. \cite{Hannun2014DeepSS} proposed data augmentation via adding additional noise to speech, reducing WER by 21.3\% relative on their self-constructed 100 sentence evaluation set. 
%However, in this approach, the essential information for the training process can be covered by noise. When the most important speech regions are severely distorted by noise, this technique can assign transcription to noise which led to a faulty training data. 

Recently, data augmentation techniques have been introduced that utilize importance or saliency maps. There are many methods to predict importance and saliency maps, e.g., \cite{itti1998model,Harel2006,jetley2016end,Kummerer_2017_ICCV,hou2007saliency,pan2017salgan,kim2017bubbleview,spille2017listening,Trinh2020,trinh2018bubble,9271908}, but few previous studies have investigated applications of such maps. In the visual domain, a recent work \cite{gong2021keepaugment} used saliency maps for data augmentation. Instead of using noise, the authors cut random rectangles out of an image if the sum of the importance scores of all the pixels inside the rectangle was smaller than a threshold. In speech, \cite{do2018weighting} used a bottom-up approach to predicting auditory saliency maps to improve ASR performance. They used Gabor filters to extract intensity and contrast in time and frequency to find the saliency maps. This saliency map is then multiplied with the spectrogram, resulting in a weighted spectrogram, from which features are extracted for ASR. This approach achieved a 5.3\% relative WER reduction compared to a baseline that did not use importance maps. 

We introduced a top-down adversarial approach to predicting importance maps in \cite{trinh2018bubble,kavaki20_interspeech}. The current paper builds upon those approaches to introduce a method of using our top-down importance maps for data augmentation in speech command recognition. In contrast to \cite{do2018weighting}, we use a top-down approach to identify the regions that are important for recognizing the specific production of the specific words in a given utterance. Furthermore, these regions are directly related to the speech recognition task, which is different from bottom-up approaches, which produce the same prediction regardless of the task. For instance, a bottom up approach using intensity filters might predict that a spectrogram area containing loud noise is important for the speech recognition task. 
%Second, our method keeps the most essential information for training purpose and added noise only to unimportant regions in the spectrogram instead of adding noise everywhere. 
%On the Google speech commands utterance classification dataset \cite{warden2018speech} this approach achieves a 14.8\% and 23.7\% relative error rate reduction compared to conventional noise augmentation and no augmentation, respectively. 

%%%%%%%%%%%%
\begin{figure*}%[!hbtp]
\centering
\begin{adjustbox}{width=\textwidth}
\begin{tikzpicture}[every node/.style={font=\footnotesize}]
\node (word2_pic)[inner sep=0pt]  
    {\includegraphics[width=0.1\textwidth]{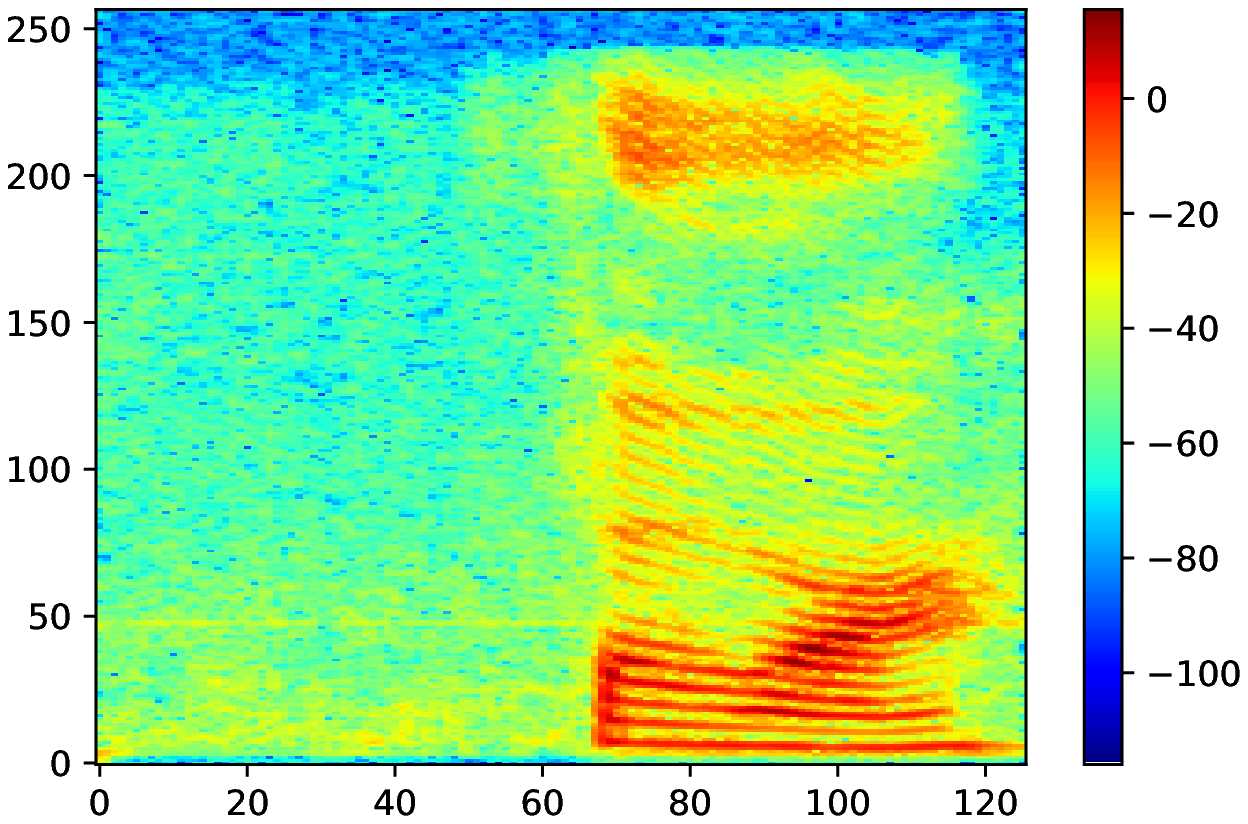}};
\node (dot1) [c,right of = word2_pic, xshift = 0.5 cm, fill = black,minimum size=0.5em]{};    
\node (text) [above of= word2_pic,yshift=-0.2cm] {Clean speech $S$};
\node (gen) [process, right of =dot1,xshift= 0.7 cm] {Generator G};
\node (mask_pic) [inner sep=0pt,right of= gen, xshift=1.5 cm]  
    {\includegraphics[width=0.1\textwidth]{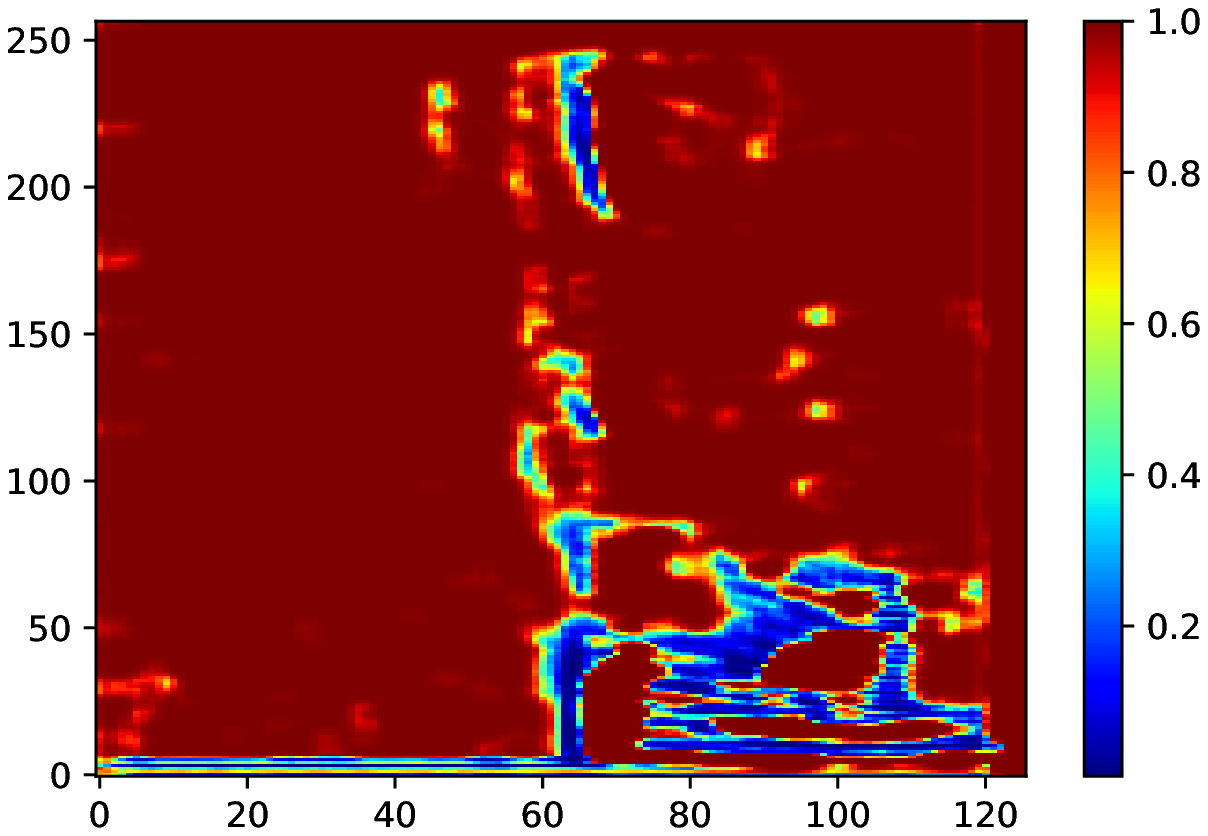}};

\node (multi1) [c, right of= mask_pic, xshift=1 cm] {$\odot$};
\node (add2) [c, right of= multi1, xshift=5 cm] {$+$};

%\node (G1) [tria,below of =add2, xshift= -1 cm,yshift= -0.2 cm] {a};
\node (G2) [tria_r,below of =multi1, xshift= 1 cm,yshift= -0.2 cm] {A};
%\node (noise) [process,right of=G2,xshift= 1 cm] {Noise};
 \node[inner sep=0pt,right of=G2, xshift=1 cm] (noise)
    {\includegraphics[width=0.1\textwidth]{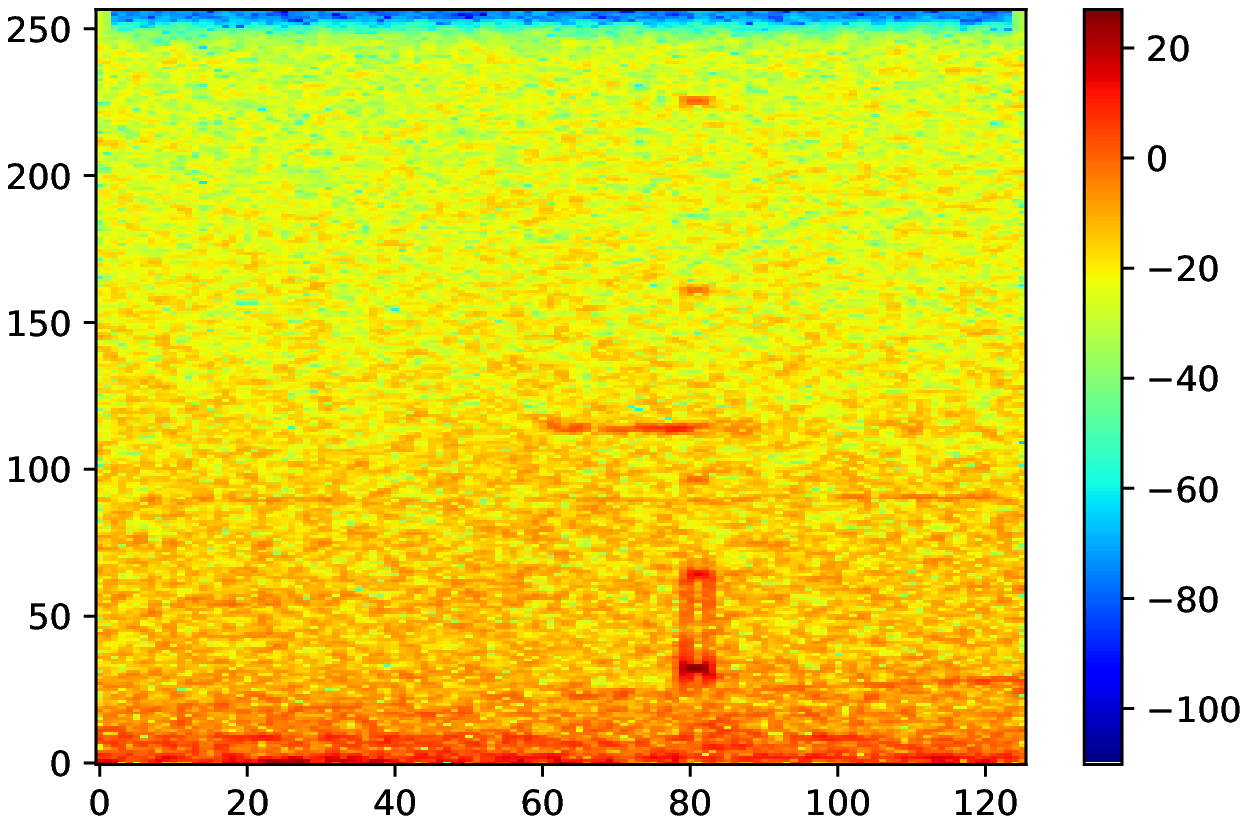}};
\node (text) [above of= noise,yshift=-0.2cm] {Noise N};
\node (text) [above of= mask_pic,yshift=-0.2cm] {Importance map $M_{\theta}$};
 \node[inner sep=0pt,above of= add2, xshift=2.2 cm, yshift= -0.2 cm] (noisy_pic) 
    {\includegraphics[width=0.1\textwidth]{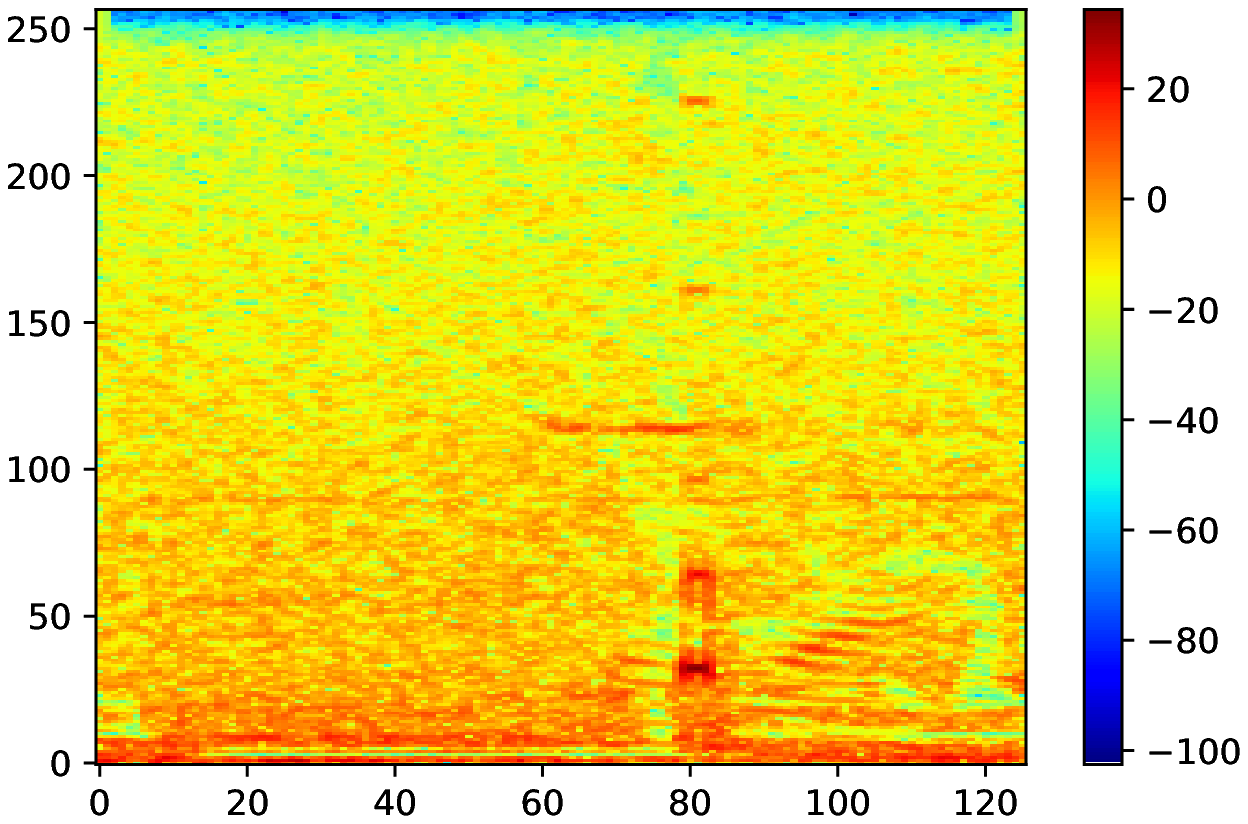}};
\node (text) [above of= noisy_pic,yshift=-0.2cm] {Noisy mixtures};
\node (SCR) [process, right of= add2, xshift=4 cm] {Speech Command Recognizer};
\draw (dot1) -- ($(dot1)+(0,1.1)$); 
\draw [arrow] ($(dot1)+(0,1.1)$) -| (add2); 
%\draw [arrow] (noise) -- (G1);
\draw [arrow] (noise) -- (G2);
\draw (word2_pic) -- (dot1);
\draw [arrow] (dot1) -- (gen);
\draw [arrow] (gen) -- (mask_pic);
\draw [arrow] (mask_pic) -- (multi1);
\draw [arrow] (multi1) -- (add2);
\draw [arrow] (add2) -- (SCR);
\draw[arrow,rounded corners=5pt] (G2) - | (multi1);
%\draw[arrow,rounded corners=5pt] (G1) - | (add2);
%\draw[arrow,rounded corners=5pt] (word2_pic) - | (add2);
%\draw[arrow,rounded corners=5pt] (G2) - | (add1);
\end{tikzpicture}
\end{adjustbox}
\caption{ImportantAug scheme. The mask generator's task is to output an importance map (mask) for an utterance with maximal noise while interfering with recognition of the recognizer as little as possible. The mask is point-wise multiplied ($\odot$) with the scaled noise and added to the clean speech. The mask contains values close to 0 at important points and values close to 1 at unimportant points.}
\label{fig:impAugArch}
\end{figure*}
%%%%%%%%%%%%%%%%%%%%%%%%%%%%%%%%%%%

In section 2, we discuss our ImportantAug\footnote[1]{The code is available at https://github.com/tvanh512/importantAug} method, where we first identify the importance maps and then utilize them to augment the data. In
section 3, we present our experimental setup with details about the data, hyperparameter settings, and experiments. The results on clean, in domain
noisy, and out-of-domain noisy test sets are illustrated in section 4.

\section{Method}

The proposed network has a speech command recognizer and a mask generator, as illustrated in Figure \ref{fig:impAugArch}. The speech command recognizer's task is to classify the input utterances into the correct classes. The mask generator's task is to add as much noise as possible to utterances without harming the performance of the recognizer. This has the effect of generating importance maps, which are utilized for data augmentation. 

Our networks are trained in two stages. In the first stage, we train the generator so that it can output importance maps (masks). We load a recognizer that is pre-trained on clean speech. Then, we freeze the recognizer and train only the mask generator. The generator receives clean speech as input and outputs a mask. This mask is multiplied with the noise and then added to the clean speech, resulting in a noisy utterance. The recognizer receives this noisy speech as input and predicts a class. Note that in the Google Speech Commands (GSC) dataset \cite{warden2018speech}, each utterance is at most 1s long and only contains a single word in the presence of noise. 
% although utterances can vary in length. 
Thus this is a speech classification task as opposed to a full speech recognition task.

We designed the loss function for our network to encourage the mask to maximize the amount of noise while the speech recognizer maintains good performance. This loss function therefore forces the generator to output a mask with less noise in regions that are important to the recognizer, and with more noise in regions that are unimportant to the recognizer.

%-------------------
%We . We have the generator which can output importance maps (e.g. Figure \ref{fig:Important_aug_process}.b) as a result from stage 1.

In the second stage, we freeze the generator and train only the speech command recognizer. We aim to create additional data to train the recognizer. To create additional data, noise is added to the unimportant regions of the clean speech. Less or no noise is added to the important regions. 

%In addition, we introduce a second augmentation by setting a random horizontal and vertical stripe in the mask to zero. This is necessary because XXXXXX. The frequency bins from $f_{start}, f_{start} + f_{width}$ have value 0. $f_width$ are uniformly samples from 0 to the parameter $f_{mask}$. $f_{start}$ are sample from $0,f_{bins}-f_{width}$. The time frames also masked with 0 in the same way, but with parameter $t_{width}$. Then, the importance maps with stripes (e.g. Figure \ref{fig:Important_aug_process}.c) will be multiplied with the noise (e.g. Figure \ref{fig:Important_aug_process}.d) and added to the speech, resulting in masked noisy speech (e.g., Figure \ref{fig:Important_aug_process}.e). The speech information is maintained inside the horizontal and vertical stripes, which is different from other methods, such as SpecAug, where the speech information inside the stripes are removed. 

Denote $S(f,t) $ and $N(f,t)$ as the complex spectrograms of the speech and noise, respectively, where $f$ is the frequency index and $t$ is the time index. These spectrograms are created by applying the short time Fourier transform (STFT) to the time domain signal $s(t)$ of the speech and $n(t)$ of the noise. The generator $G$ with parameters $\theta$ takes $\tilde{S}(f,t) = 20 \log_{10}|S(f,t)|$ as input and predicts a mask $M_{\theta}(f,t)$ with the same shape as $\tilde{S}(f,t)$
%%%
\begin{align}
M_{\theta}(f,t) = G(\tilde{S}(f,t); \theta) \in [0,1]^{F \times T}
\end{align}

An additional augmentation shifts the mask slightly in time or frequency to further increase variability in the training data for the recognizer. The mask output by the generator, $M_{\theta}$, is rolled along the frequency and time dimension 
%%%
\begin{align}
M_{\theta r} = r(M_{\theta};\delta)
\end{align}
%%%
where r is the roll operator (we use torch.roll) and $\delta$ is the number of time frames or frequency bins by which the elements of the mask are shifted. $\delta$ is drawn uniformly at random from the interval $(-D,D)$. 
Furthermore, to create additional variation, with probability $0.5$, the mask $M_{\theta r}$ is replaced by a mask of all 1's. Denote whichever mask is selected as $M$.
% or all 1s depends on a random variable $v$ that is drawn randomly in a uniform distribution over two entries 0,1. 
% \begin{align}
% M_{\theta r v}=
% \begin{cases} M_{\theta r}  & \text{if $v =0$} \\
% 1 &  \text{if $v=1$}
% \end{cases}
% \label{eq:random_v}
% \end{align}
This rolling augmentation is only used when re-training the recognizer using the predicted importance maps and not when training the mask generator itself.

%The mask is rolled in order to maintain selected subset of the important regions of the speech to be noise free when adding. This technique inherent and resemble the power of boosting method in machine learning because it chooses different subsets of (important) features to train the network at different iterations, thus preventing the discriminator from overfitting. We choose the $\Delta$ parameters such that the mask is shifted nearby the original speech, thus the selected subsets includes parts of important regions. Thus the networks still have enough useful information for training purpose.   

This mask is then applied point-wise to a noise instance $N$, scaled by gain $A$. 
The gain A is adjusted each training batch such that the signal to noise ratio is maintain at a target value
\begin{align}
A=\sqrt{\frac{\sum_{b,t,f}|S_{btf}|^2}{10^{v/10} \sum_{b,t,f} |N_{btf}|^2 }},
\end{align}
where $v$ is the target SNR expressed in decibels, and b, t, f denoted the batch, time, and frequency dimensions respectively. The resulting masked-scaled noise $A N \odot M$ (where $\odot$ denotes point-wise multiplication) is added to the clean speech $S$. The resulting noisy mixture is input to the speech command recognizer $R$, which predicts the probability of the class $\hat{y}$
%%%
\begin{align}
    \hat{y} = R(S + A N \odot M).
\end{align}
%%%

% When the model is in stage one, M is denoted $M_{\theta r v}$ and when the model is in stage two, M is denoted $M_{\theta r}$ because we only apply roll operation when we trained the classifier in stage 2. In stage 1, we train the generator to output the correct mask, so we do not apply any operation. 

The model is trained to minimize
%---------------------
\begin{align}
  \mathcal{L}(\theta) &= \lambda_r \mathcal{L}_{\textrm{R}}(y, \hat{y}) 
  - \frac{\lambda_e}{TF}\sum_{f,t} \log M \nonumber \\ 
  &+ \frac{\lambda_f}{TF}\sum_{f,t} |\Delta_f M| 
  + \frac{\lambda_t}{TF}\sum_{f,t} |\Delta_t M|. \label{eq:bcnLoss}
\end{align}
%---------------------
where $\mathcal{L}_{\textrm{R}}$ is the loss of the speech recognizer, $\Delta_f$ is the difference operation along frequency, $\Delta_t$ is the difference operation along time, and $\lambda_r, \lambda_e, \lambda_f,$ and $\lambda_t$ are weights set as hyperparameters of the model. The recognizer loss is the cross entropy between the prediction $\hat{y}$ and the ground truth label $y$. This loss forces the recognizer to keep high accuracy on predicting the correct class. The $-\sum_{f,t} \log M$ term forces the mask's value to be close to one, thus maximizing the amount of noise added. The terms associated with $\lambda_f$ and $\lambda_t$ encourage the mask to smooth in frequency and time. 

%---------------------
\section{Experimental setup}
\subsection{Dataset}
We use the Google Speech Commands (GSC) dataset version 2 \cite{warden2018speech} for our experiments. This dataset includes 105,829 single-word utterances of 35 unique words. %Some examples of the classes are ``up," ``down", ``left," and ``right,". 
Many utterances include noise or other distortions. The models were trained on the training set and evaluated on the test set. The development set was used for early stopping. 

We also employ additional noise from the MUSAN dataset \cite{snyder2015musan} to augment the speech from the GSC dataset. The recordings in MUSAN have different lengths, so we only used the first second from each recording and exclude any recordings shorter than one second as the speech utterances are restricted to be at most one second long. There are 877 noisy files after filtering out short utterances. We randomly choose 702 files (80\%) for training. We mix the remaining 175 files with the utterances from the GSC test set, creating a new noisy test set that we call GSC-MUSAN. 

To evaluate our trained model on out-of-domain noisy environments, we also create another test set. First, we select a file ``HOME-LIVINGB-1.wav", which contains 40 minutes of noise recording in the living room environment from the QUT corpus \cite{dean2015qut}. We then resample this file from 48 to 16~kHz, the same rate as the GSC utterances and choose random sections in this noisy file to mix with the utterances in the GSC test set. We call this dataset GSC-QUT.

%\begin{table}
%  \caption{Recognizer error rate (\%) on the Google Speech Command v2 development set with conventional noise augmentation at different SNRs}
%  \label{tab:importantAug_result1}
%  \begin{center}
%  \begin{tabular}{cc@{$\qquad$}cc}
%  \toprule
%  SNR & Dev error & SNR & Dev error \\
% \midrule
%  $\infty$ & 7.74 & 15 & 6.09 \\  
%  40 & 6.27 & 10 & \textbf{5.87} \\
%  35 & 6.36 & 5 & 6.30 \\
%  30 & 6.37 & 0 & 6.26 \\
%  25 & 7.90 & $-5$ & 6.54 \\
%  20 & 6.17 &  $-10$ & 6.47 \\
%  \bottomrule
%    \end{tabular}
%  \end{center}
% \end{table}

\begin{table}
  \caption{Recognizer error rate (\%) on the Google Speech Command v2 (GSC) development set with conventional noise augmentation at different SNRs}
  \label{tab:importantAug_result1}
  
  \begin{center}
  %\scriptsize
  \begin{tabular}{cc@{$\qquad\quad$}cc}
  \toprule
  SNR & Dev  & SNR & Dev  \\
  \midrule
  $\infty$ & 7.74 & 15 & \textbf{5.83} \\  
  40 & 6.39 & 10 & 6.11 \\
  35 & 7.65 & 5 & 6.00 \\
  30 & 6.10 & 0 & 5.97 \\
  25 & 6.19 & $-5$ & 6.24 \\
  20 & 6.22 &  $-10$ & 6.16 \\
  \bottomrule
    \end{tabular}
  \end{center}
\end{table}

\subsection{Experiments}

We compare our proposed method against two other methods. In the first method (baseline), we train a recognizer that does not utilize any data augmentation . It is trained on the GSC training set and selected using early stopping on the development set. All other methods are trained by initializing their parameters to those of this pre-trained baseline recognizer. 
%In \textbf{experiment 2}, we compare our proposed method against a conventional noise augmentation technique that applies noise directly to the speech without importance maps.
In the second method, we utilize a conventional noise augmentation technique that treats all time-frequency points as equally important and applies noise directly to the speech without importance maps (S + AN). We perform an experiment to identify the best single signal to noise ratio (SNR) to use, comparing those ranging from $-10$~dB to 40~dB in steps of 5~dB. We also evaluate $\infty$~dB by training on clean data. 

In our proposed method, ImportantAug, we performed the two-stage training as described above. First, we load and freeze the recognizer from the baseline and train the generator. Then, we freeze the generator and train the recognizer. The noise from the MUSAN dataset was multiplied with the rolled importance maps and added to the speech. In addition, we perform an ablation study by evaluating the recognizer performance when we remove the importance map from the proposed approach, by setting the mask to be all 1's, which we call the ``Null ImportantAug" condition. In this case, no region is more important than other regions and the noise is added directly to the speech. 
We evaluate the baseline (no augmentation), conventional noise augmentation, ImportantAug and Null ImportantAug on the standard GSC test set, GSC-MUSAN and GSC-QUT noisy test sets. 

%-------------------
\begin{figure*}
  \begin{center}
    \scriptsize
    \begin{tabular}{ccccc}
      \includegraphics[width=0.18\textwidth]{figures_importantAug/clean_speech.eps} &
      \includegraphics[width=0.18\textwidth]{figures_importantAug/mask.eps} &
      \includegraphics[width=0.18\textwidth]{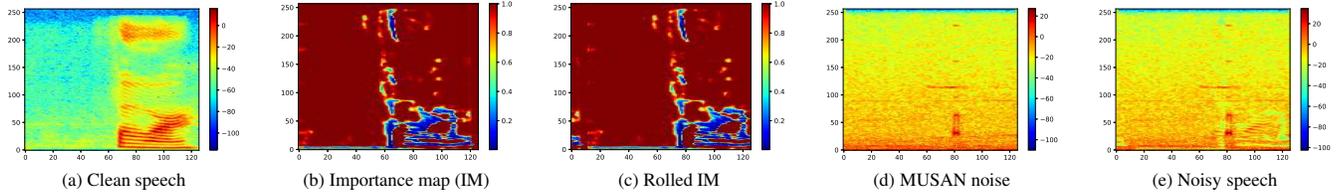} &
      \includegraphics[width=0.18\textwidth]{figures_importantAug/noise.eps} &
      \includegraphics[width=0.18\textwidth]{figures_importantAug/x2.eps} \\
      (a) Clean speech &
      (b) Importance map (IM) &
      (c) Rolled IM &
      (d) MUSAN noise &
      (e) Noisy speech \\
    \end{tabular}
  \end{center}
  \caption{
  (a) Clean utterance from Google Speech Commands dataset. (b) Importance map (blue areas) from the generator. (c) Rolled importance map. (d) MUSAN noise. (e) Noisy speech created by multiplying the noise from (d) with the mask from (c) and adding clean speech from (a) }
  \label{fig:Important_aug_process}
\end{figure*}

In addition to using continuous-valued importance maps, we also experimented with binarizing the importance maps . We considered the $q$\% of time-frequency points with the lowest value in the continuous-valued importance map as being important and did not add any noise to them. The other $100-q$\% of the points were considered unimportant and noise was added to them. In this experiment, the mask was not replaced by an all 1's mask at all.
%we fix $v=0$ in \eqref{eq:random_v} because the mask is all 1s with $v=1$. This leads to $q=0$ and we do not want q to receive both zero and non-zero value at different iterations. 
%he q is an experimental number and is chosen in a way that minimizes the recognizer error rate.

\subsection{Hyperparameter settings}

The signal was sampled at 16~kHz with a window length of 512 and a hop length of 128 samples, leading to a spectrogram with 257 frequency bins and 126 time frames for a 1~s utterance. In all experiments, we use the same default setting for the speech command recognizer, which is a neural network with 5 layers. Each layer has a 1D depth-wise and 1D point-wise convolution \cite{chollet2017xception, Somshubra20}, followed by SELU activation \cite{klambauer2017self}. The depth-wise convolution has a kernel size of $9\times 9$ (281.25 Hz x 96 ms), a stride value of 1, a dilation value of 1; and its inputs and outputs are both 257 channels. The point-wise convolution consists of a kernel of size $1 \times 1$ and also has inputs and outputs for size 257. 

The generator is a neural network with 4 layers, where each layer is a 2D convolutional network. The first layer takes one channel in and outputs 2 channels. The second and third layers have 2 channels in their input and output. The last layer has 2 channels of input and one of output. All the layers have a kernel size of $5 \times 5$ (156.25 Hz x 64ms), a stride value of 1, a dilation value of 1 and a padding so that the output has the same height and width as the input. 

In the proposed ImportantAug method, we selected hyperparameters $\lambda_r = 1$, $\lambda_e=\lambda_f=\lambda_t=3$, $v=-12.5$~dB. First, the weights $\lambda_r, \lambda_e, \lambda_f, \lambda_t$, and $v$ were manually adjusted based on a very small number of settings so that the speech command recognizer performed well and the mask values were closer to all 1's on the development set. Then we chose $D$, the maximum number of time frames or frequency bins by which the elements of the mask are shifted to be 30, equivalent to 937.5 Hz and 264 ms. This was selected to keep the mask from shifting too far from the original position.

All the models are trained with the Adam optimizer with an initial learning rate of $0.001$, which is decayed  by half every 20 epochs and a batch size of 256. The models are trained for 200 epochs with early stopping on the development set loss with a patience value of 30.

%%%%%%%%%%%%%%%%%%%%%%%%%%%%%%%%%%%%%%%%%%%%%%%%%%
\section{Results}
Table~\ref{tab:importantAug_result1} shows the error rate on the development set for the conventional augmentation method with different signal to noise ratios. We can see that adding too much noise leads to a high error rate, for example, SNRs -10 and -5~dB have error rates 6.16\% and 6.24\%, respectively on the development set. Adding too little noise is also not optimal, for instance, SNRs 40 and 35~dB have error rate 6.39\% and 7.65\% on the development set. Using no noise at all does not provide good performance, with an error rate 7.74\%. However, adding the right amount of noise is beneficial for the recognizer as it balances variation in the training data with speech fidelity. As shown in Table~\ref{tab:importantAug_result1}, the best error rate (5.83\%) is with an SNR of 15~dB. The model trained with SNR 15~dB has the best performance on the development set, so we choose this model to evaluate on the test set and compare with other approaches in Table~\ref{tab:importantAug_result2}. 

\begin{table}
  \caption{Recognizer error rate (\%) with various augmentation approaches on GSC test set} \label{tab:importantAug_result2}
  \begin{center}
    % \scriptsize
    \begin{tabular}{lcc}
    \toprule
      Augmentation method & Initial SNR (dB) & Error \\
     \midrule
     No augmentation & $\infty$  & 6.70 \\
      Conventional noise augmentation & 15.0 & 6.52  \\
      ImportantAug & -12.5 & \textbf{5.00} \\
      Null ImportantAug & -12.5 & 6.12 \\
      \bottomrule
    \end{tabular}
  \end{center}
\end{table}

Table~\ref{tab:importantAug_result2} shows the results on the standard GSC test set. The baseline speech command recognizer has an error rate of 6.70\%. The conventional noise augmentation method produces a model with an error rate of 6.52\%. Our proposed method has the best error rate at 5.00\%, which is a 25.4\% relative improvement over the no augmentation baseline and 23.3\% relative improvement over the conventional noise augmentation method. We also perform an ablation study with the Null ImportantAug method by using a ``mask'' that is all 1's, which leads to an error rate of 6.12\%. Null ImportantAug is similar to the traditional NoiseAug because it does not utilize importance maps. The difference is that Null ImportantAug is trained with the same SNR as ImportantAug (-12.5 dB), while the traditional NoiseAug uses the SNR chosen based on the
performance on the development set of 15 dB. The error rate with and without important maps are 5\% and 6.12\% respectively,  thus the importance map is necessary for the observed performance gains.

\begin{table}
  \caption{Recognizer error rate (\%) of augmentations on in-domain noise test set (GSC-MUSAN) as a function of test SNR.}% This test set is created by mixing utterances in GSC test set with random MUSAN noise utterances (not use in training) at different SNRs from $-12.5$ to $40$} 
  \label{tab:importantAug_result_MUSAN}
  \begin{center}
  \scriptsize
    \begin{tabular}{lrrrrrrr}
    \toprule
    & \multicolumn{7}{c}{Test SNR} \\
     Method &  $-12.5$ &$-10$ &0 & 10 & 20 & 30 &40 \\
     \midrule
     No aug. (baseline) & 77.6 & 72.7 & 45.2 & 21.0 & 11.5 & 8.4 & 7.3 \\
       Noise aug.~(SNR 15) & 65.8 & 57.7 & 26.3 & 10.8 & 7.3 & 6.6 & 6.4  \\
        ImportantAug & \textbf{43.5} & \textbf{35.0} & \textbf{13.3} & \textbf{7.4} & \textbf{5.7} & \textbf{5.2} & \textbf{5.1}   \\
        Null ImportantAug & 45.2 & 37.0 & 15.0 & 8.5 & 6.9 & 6.2 & 6.0  \\
      \bottomrule
    \end{tabular}
  \end{center}
\end{table}

Table~\ref{tab:importantAug_result_MUSAN} shows the results on the GSC-MUSAN test set. We could observe that the proposed method ImportantAug achieve the best result in all SNR range. For example, the ImportantAug achieve 13.3\% error rate at 0~dB, which is around one-third of the error rate of the baseline 45.2\% and a half of the conventional augmentation method. We also observe that the error rates are going up if we remove the importance map (IM) when comparing row 3 and row 4 of Table~\ref{tab:importantAug_result_MUSAN}. For example, at SNR 0~dB, the error rate going up from 13.3\% to 15\% if we remove the IM and train with only the noise.  
%-------------------------------
%---------------------------

Table~\ref{tab:importantAug_result_QUT} shows the results on the GSC-QUT test set, which is out-of-domain noise test set because the models are trained with MUSAN noise, not with QUT noise. Here, we observe the same trend when the ImportantAug outperforms the baseline, the conventional augmentation method.

\begin{table}
  \caption{Recognizer error rate (\%) of augmentations on out-of-domain noise test set (GSC-QUT) as a function of test SNR.} %This test set is created by mixing utterances in GSC test set with random QUT noise utterances at different SNRs from $-12.5$ to $40$} 
  \label{tab:importantAug_result_QUT}
  \begin{center}
     \scriptsize
    \begin{tabular}{lrrrrrrr}
    \toprule
    & \multicolumn{7}{c}{Test SNR} \\
     Method &  $-12.5$ &$-10$ &0 & 10 & 20 & 30 &40 \\
     \midrule
     No aug. (baseline) & 90.9 & 87.3 & 55.8 & 20.8 & 9.6 & 7.4 & 7.0  \\
       Noise aug.~(SNR 15) & 89.0 & 83.5 & 42.0 & 12.9 & 7.3 & 6.5 & 6.2   \\
        ImportantAug & \textbf{72.0} & \textbf{61.3} & \textbf{23.5} & \textbf{8.9} & \textbf{5.8} & \textbf{5.1} & \textbf{4.8}  \\
        Null ImportantAug & 72.3 & 61.6 & 24.8 & 10.0 & 6.8 & 6.1 & 6.0  \\
      \bottomrule
    \end{tabular}
  \end{center}
\end{table}
Figure~\ref{fig:Important_aug_process}.b shows an example of an importance map of an utterance of the word ``four" in the GSC dataset. The importance map includes the fundamental frequency, the harmonics, and the outer border shape of the speech. These regions are predicted to be necessary for the speech command recognizer to identify this specific utterance. Thus, keeping these regions clean and adding noise outside of them makes the data more diverse while not affecting the recognition.

% \begin{table}
%  \caption{Recognizer error rate (\%) with binarized ImportantAug using different important region ratios, $q$.}
%  \label{tab:importantAug_result3}
%   \begin{center}
%     %\scriptsize
%     \begin{tabular}{ccc}
%     \toprule
%      $q$ (\% )  & Dev error & Test error \\ 
%      \midrule
%      100 & 6.50 & 6.70 \\
%      70 & 6.37 & 6.82 \\
%      50 & 7.12 & 7.44 \\
%      40 & 6.95 & 7.32 \\
%      20 & 6.67 & 7.01 \\
%      10 & 6.21 & 6.43 \\
%     4 & \textbf{6.01} & \textbf{6.26} \\
%     1 & 6.05 & 6.28 \\
%      0 & 6.58 & 6.52 \\
%      \bottomrule
%   \end{tabular}
%  \end{center}
% \end{table}

\begin{table}
 \caption{Recognizer error rate (\%) with binarized ImportantAug using different important region ratios, $q$ on the original GSC test set.}
 \label{tab:importantAug_result3}
  \begin{center}
    %\scriptsize
    \begin{tabular}{ccc}
    \toprule
     $q$ (\%)  & Dev & Test  \\ 
     \midrule
     %100 & 6.50 & 6.70 \\
     70 &  5.42 & 5.64 \\  
     50 &  5.49 &  5.92 \\
     40 &  5.19 &  5.71 \\ 
     20 &  5.17  & 5.15   \\
     10 & \textbf{5.00} & 5.43 \\
      5 &  5.09 &  4.92 \\ 
      1 & 5.12 & 4.94\\
      0 & 6.03 &  6.12\\
     \bottomrule
  \end{tabular}
 \end{center}
\end{table}

Table~\ref{tab:importantAug_result3} shows the error rate on the development and test set for the binary ImportantAug method with different important region ratios. In this experiment, we consider the quantile $q$\% of the regions that have lowest mask value to be important. The best result is achieved on the development set by choosing 10\% of points to be important, which provides a 11.3\% relative error reduction on the test set compared to not multiplying the noise with the importance map ($q=0$). Thus only a very small proportion of points need to be preserved in this way to enhance the data augmentation performance.
%using the raw continuous mask%

\section{Conclusion}
In conclusion, we have demonstrated a data augmentation agent that improves a speech command recognizer. Our proposed ImportantAug method produced a 25.4\% relative error rate reduction compared to the non-augmentation method and and 23.3\% relative reduction compared to the conventional noise augmentation method. Taken together, this work shows that importance maps can be estimated accurately enough to be helpful for data augmentation, providing one of the first such demonstrations, especially for speech. In the future, we will extend this framework by replacing the speech command recognizer with a full large vocabulary continuous speech recognizer and we will deploy different methods to identify the importance map and use the map to augment the speech data, such as those based on human responses. The proposed method could also be used in computer vision tasks, such as image recognition by predicting importance maps for images. 

\section{Acknowledgements}

This material is based upon work supported by the National Science Foundation (NSF) under Grant IIS-1750383. Any opinions, findings, and conclusions or recommendations expressed in this material are those of the author(s) and do not necessarily reflect the views of the NSF.

\balance
\bibliographystyle{IEEEbib}
\bibliography{strings,refs} 

\end{document}